\input{aipcheck}

\documentclass[
    ,final            
  ]
  {aipproc}

\layoutstyle{6x9}

\begin{document}

\title{A Bayesian Analysis of HAT-P-7b Using the EXONEST Algorithm}

\classification{
                \texttt{97.82.Cp, 97.75.Wx}}

\author{Ben Placek}{
  address={Department of Physics, University at Albany (SUNY), Albany NY, USA}
}

\author{Kevin H. Knuth}{
  address={Department of Physics, University at Albany (SUNY), Albany NY, USA}
 ,altaddress={Department of Informatics, University at Albany (SUNY), Albany NY, USA}
}

\begin{abstract}
The study of exoplanets (planets orbiting other stars) is revolutionizing the way we view our universe.  High-precision photometric data provided by the Kepler Space Telescope (Kepler) enables not only the detection of such planets, but also their characterization.  This presents a unique opportunity to apply Bayesian methods to better characterize the multitude of previously confirmed exoplanets.  This paper focuses on applying the EXONEST algorithm to characterize the transiting short-period-hot-Jupiter, HAT-P-7b.  EXONEST evaluates a suite of exoplanet photometric models by applying Bayesian Model Selection, which is implemented with the MultiNest algorithm.  These models take into account planetary effects, such as reflected light and thermal emissions, as well as the effect of the planetary motion on the host star, such as Doppler beaming, or boosting, of light from the reflex motion of the host star, and photometric variations due to the planet-induced ellipsoidal shape of the host star.  By calculating model evidences, one can determine which model best describes the observed data, thus identifying which effects dominate the planetary system.  Presented are parameter estimates and model evidences for HAT-P-7b.
\end{abstract}

\maketitle


\section{Introduction}
The Kepler Space Telescope (Kepler) was designed to continously moniter the brightness of stars in its field of view, which allows astronomers to collect long duration photometric time-series data on approximately 150,000 main-sequence stars.  Indications of the presence of exoplanets in the Kepler light curves stem from two periodic dimming events called transits, which are caused by the planet(s) crossing between the star and the observer's line of sight and thus eclipsing it's host star and the planet passing behind the stellar disk so that any reflected or thermally-emitted light from the planet are blocked by the host star.  These dimming events are exceedingly rare as they require the planet to be within a very small range of orbital inclinations.  Fortunately, in addition to transits there are four other physical mechanisms that induce periodic photometric variations that do not disappear if the planet is non-transiting, which describes the majority of the extant planet population.  The first two effects are the reflection of the star-light off of the planetary atmosphere (or surface), and the thermally emitted light from the day and night sides of the planet.  The last two effects are caused by the planet's effect on it's host star. Doppler beaming is a relativistic effect that causes stellar-emitted light to be `beamed' along the direction of it's motion.  Since planets and stars co-orbit their center of mass, the star will at times move toward and away from a distant observer, thus causing the emitted light to be more intense when the star is moving toward the observer and less intense when the star is moving away from the observer.  Finally, the proximity of the planet to its host star induces tidal bulges on the stellar surface causing the star to take on an ellipsoidal shape.  As the planet orbits the host star, the ellipsoid rotates, which induces changes in brightness due to the changing apparent cross-sectional area of the star as well as gravity darkening effects.  It has been shown that one can detect and characterize exoplanets by analyzing these four photometric effects alone  \citep{Placek+etal:2014,Esteves+etal:2013,Shporer+etal:2011,Faigler&Mazeh:2011,Faigler+etal:2013,Faigler&Mazeh:2014}.

The EXONEST algorithm developed by \citep{Placek+etal:2014} utilizes Bayesian model selection in order to characterize both transiting, and non-transiting exoplanets.  By calculating the Bayesian evidence, one can test a suite of models each taking a different subset of photometric effects into account, in order to determine which model best describes a given dataset.  This methodology was shown to accurately characterize the transiting planet Kepler-13b (previously KOI-13b) using out-of-transit data \citep{Placek+etal:2014}.  Including the transits and secondary eclipses is expected to yield more accurate parameter estimates since these dimming events hold information about the orbital inclination, planetary radius, and dayside temperature. 

This work aims to add to the work done previously by Placek et al. \citep{Placek+etal:2014} by including transits in the photometric models.  EXONEST is applied to the light curve of HAT-P-7b, a transiting hot-Jupiter that orbits it's host star in just $2.2025$ days.  Bayesian model selection indicates that HAT-P-7b is in a circular orbit.  Relevant parameter estimates for HAT-P-7b are also obtained as a by-product of EXONEST.

\section{The EXONEST Algorithm}

The EXONEST Algorithm relies on Bayesian model selection to characterize exoplanetary signals.  In order to perform model testing, a number of inputs are required.  First, is the forward model, which includes all of the desired photometric effects (Reflection, Thermal Emissions, etc.).  Next are the prior probability assignments.  These probability distributions quantify one's prior state of knowledge before analyzing any data.  Last is the assignment of the likelihood function, which quantifies the degree to which one expects the data to agree with the photometric time-series predicted by the forward model.  With the forward model, the prior probabilities, and the likelihood function, EXONEST uses the MultiNest algorithm \citep{Feroz&Hobson:2008,Feroz+etal:2009,Feroz+etal:2013} to calculate log-evidence estimates and parameter estimates.

\subsection{The Photometric Effects}
Each of the photometric effects described in the Introduction causes very specific signals in the observed light curve. This section will briefly outline the model for reflected light, thermal emission, Doppler boosting, and ellipsoidal variations and describe how each affects the overall light curve signal.  For a comprehensive description of these effects see Placek et al. \citep{Placek+etal:2014}.  The reflected component of the observed flux is given by
\begin{equation}
\frac{F_R(t)}{F_\star} = \frac{A_g}{2} \frac{{R_p}^2}{r(t)^2} \left(1 + \cos \theta(t) \right).
\end{equation}
where $A_g$ is the geometric albedo, which is a ratio of incident to reflected light for the planet, and $R_p$ and $r(t)$ are the planetary radius and planet-star separation distance, respectively.  Last, $\theta(t)$ is the angle between the unit vector from the star to the planet, and the observer's line of sight, which can be determined by calculating the $(x,y,z)$ position of the planet as a function of time.  These flux variations are of the same frequency as the orbit.  

The stellar normalized thermal flux from the day and night-sides of the planet can be modeled as 
\begin{equation} \label{eq:photometric-thermal-day}
\frac{F_{Th,d}(t)}{F_\star} = \frac{1}{2}(1 + \cos\theta(t)) \left( \frac{R_p}{R_\star} \right)^2 \frac{ \int B(T_d)K(\lambda) \, d\lambda}{\int B(T_{eff}) K(\lambda) \, d\lambda}
\end{equation}
and
\begin{equation} \label{eq:photometric-thermal-night}
\frac{F_{Th,n}(t)}{F_\star} =\frac{1}{2} (1 + \cos(\theta(t) - \pi)) \left( \frac{R_p}{R_\star} \right)^2 \frac{ \int B(T_n)K(\lambda) \, d\lambda}{\int B(T_{eff}) K(\lambda) \, d\lambda},
\end{equation}
where $R_\star$ is the stellar radius, $T_d$ is the day-side temperature, and $T_{eff}$ is the effective temperature of the host star.  The function $B(T)$ is the spectral radiance of a blackbody and $K(\lambda)$ is the spectral response function \citep{VanCleve&Caldwell:2009} for the Kepler photometer, which describes how bright a source emitting at wavelength $\lambda$ appears in the Kepler CCD. 

The relativistic Doppler boosting signal can be written in the non-relativistic limit as
\begin{equation} \label{eq:photometric-boosting}
\frac{F_{boost}(t)}{F_\star} =  1 + 4\frac{V_r(t)}{c},
\end{equation}
where $V_r$ is the stellar radial velocity, and $c$ is the speed of light.  The signal produced by this effect is of the same frequency as reflected and thermal flux, but is shifted in phase by $90^o$.  This is due to the fact that Doppler boosting is a maximum when the star has a maximum radial velocity, which occurs when the planet is in it's quarter phase. 

\begin{figure}
\includegraphics[scale = 0.5]{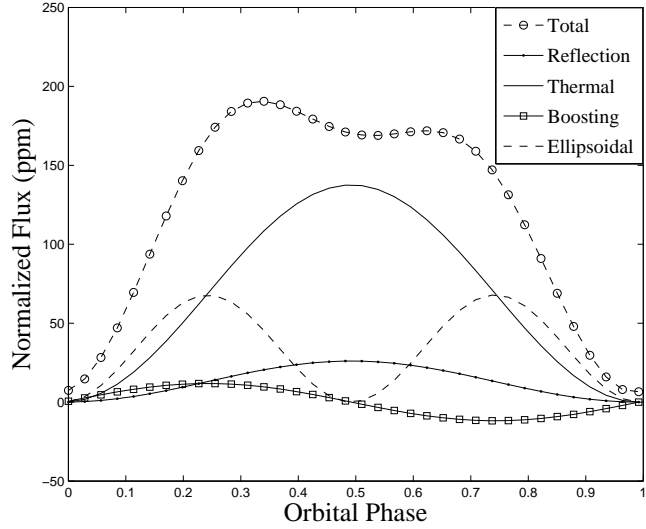}
\caption{Model-generated phase-folded light curve excluding transits and secondary eclipses.  The signal as observed through Kepler is the total photometric signal.  Model parameters used to generate this light curve were as follows: $T = 1.76d$, $e = 0$, $\omega = 0$, $M_0 =  0$, $\cos i = 0.04$, $R_\star = 2.55R_\odot$, $M_\star = 2.05 M_\odot$, $T_{eff} = 8500K$, $R_p = 1.8R_J$, $M_p = 8.35 M_J$, $T_d = 3700K$, $T_n  = 0K$, $A_g = 0.15$. }
\label{fig:model light curve}
\end{figure}

Lastly, ellipsoidal variations can be approximated by 
\begin{equation} \label{eq:photometric-ellipsoidal}
\frac{F_{ellip}(t)}{F_\star} = \beta \frac{M_p}{M_\star} \left( \frac{R_\star}{r(t)} \right)^3 [\cos^2(\omega+\nu(t)) + \sin^2(\omega+\nu(t))\cos^2i]
\end{equation}
where $\nu(t)$ is the true anomaly, $\omega$ is the argument of periastron, $e$ is the orbital eccentricity, $M_p$ and $M_\star$ are the masses of the planet and star, respectively, and $\beta$ is the stellar gravity-darkening exponent. The ellipsoidal variations create a photometric signal at twice the orbital period.  To model the observed photometric signal, one simply needs to sum each of these four signals as shown in Figure \ref{fig:model light curve}. 

\subsection{Prior Probability and Likelihood Function Assignments}
In this application of EXONEST, we assign uniform priors over a reasonable range for all model parameters.  This can be modified if more information is attained for either a single exoplanet, or exoplanets in general.  The prior probability assignments for each model parameter are shown in Table \ref{tbl:priors}.  It should be noted that in the case of the orbital inclination one must sample uniformly from a sphere, making $\cos i$ the model parameter and using a uniform prior from [0,1].  One may also choose to use a Jeffrey's prior for the signal variance which is uniform in $\log \sigma$. 

\begin{table}[h]
\begin{tabular}{l c c c c c}
  Parameter & Variable & & Interval & & Distribution \\
\hline
Eccentricity & e & & $[0,1]$ & & Uniform \\
Stellar Mass $(M_\odot)$ & $M_s$  &  & $...$ & & Known \\
Mean Anomaly & $M_0$ & & $[0,2\pi]$ &  & Uniform \\
Arg. of Periastron & $\omega$ & & $[0,2\pi]$ & & Uniform \\
Orbital Inclination &  $\cos i$ &  & $[0,1]$ & & Uniform \\
Minimum Planetary Radius $(R_J)$ & $\sqrt{A_g}R_p$ & & $[0,3]$ & & Uniform \\
Planetary Radius $(R_J)$ & $R_p$ & & $[10^{-4},4]$ & & Uniform \\
Geometric Albedo	&	$A_g$ 	& &	$[0,1]$ &	&	Uniform \\
Stellar Radius $(R_\odot)$ & $R_\star$  & & $...$ & & Known \\
Planetary Mass $(M_J)$ & $M_p $ & & $[.1,20]$ & & Uniform \\
Dayside Temperature (K) &$T_{d}$ & & $[0,5000]$ & & Uniform \\
Standard Deviation of Noise (ppm) & $\sigma$ & &$[10^{-6},10^{-4}]$ & & Uniform \\

\hline
 \end{tabular}
\caption{Prior Distributions for Model Parameters.  In the Distribution column, if a parameter is deemed known, it indicates that it is known from previous studies and kept as prior information during simulations.}
\label{tbl:priors}
\end{table}

Treating the signal variance as a model parameter, we choose a Gaussian likelihood, which depends on the sum of the square differences between the datapoints in the Kepler light curve $d_i$ taken at times $t_i$ and those generated by the forward model $F_{M}(t_i)$:
\begin{equation}
\log P(D | \theta_M, M) = -\frac{1}{2 \sigma^2} \sum_{i=1}^{N} \left( F_{M}(t_i) - d_i \right)^2 - \frac{N}{2} \log{2 \pi \sigma^2},
\end{equation}
where the summation is over the $N$ data points in the light curve. 

\subsection{The MultiNest Algorithm}
Nested sampling \citep{Sivia&Skilling:2006} is a relatively new method of calculating the Bayesian evidence.  This is performed by obtaining $N$ samples from the prior distribution, and sorting these samples according to their log-likelihoods.  At each iteration, the sample with the worst log-likelihood is discarded from the live points and replaced with a new proposal sample.  This proposal sample is determined by sampling uniformly from within the implicit log-likelihood boundary defined by the log-likelihood of the discarded sample. This process is repeated until the change in computed log-evidence falls below a threshold value. 
  While Nested Sampling is an efficient tool for calculating the Bayesian evidence, it takes some effort to sample uniformly within the implicit log-likelihood boundaries.  The MultiNest algorithm \citep{Feroz&Hobson:2008,Feroz+etal:2009,Feroz+etal:2013} clusters the live samples into ellipsoids bounded by the worst log-likelihood value. The new proposal samples are only sampled from these ellipsoids, which significantly increases the efficiency of the algorithm. 




%
\section{The HAT-P-7 System}

During its commissioning phase, Kepler confirmed the existence of a hot-jupiter planet known as HAT-P-7b \citep{Borucki+etal:2009}, which is now also known as Kepler-2b.  This planet was first photometrically and spectroscopically discovered by the HATnet project, which is a network of ground-based telescopes spanning both northern and southern hemispheres \citep{Pal+etal:2008}.  HAT-P-7b is a transiting planet in a near-circular orbit of $2.205$ days around an F-type host star with mass $M_{\star} = 1.47 M_{\odot}$ and radius $R_{\star} = 1.84 R_{\odot}$ \citep{Pal+etal:2008}.  
The light curve of HAT-P-7b presents a new challenge for EXONEST due to the small amplitude of the Doppler beaming effect and ellipsoidal variations, both of which are not clearly seen by the naked eye. 

\subsection{Model Selection and Parameter Estimation}
Bayesian model selection allows one to characterize exoplanets by means of testing a suite of models that either account for or neglect certain photometric effects in order to determine which effects are significant enough to describe the data.  In this section, we apply 18 different models to the out-of-transit data of HAT-P-7b.  Table \ref{tbl:evidences} shows the log-evidences obtained from MultiNest for each of the 18 models.  Shown in bold, there are two most favored models that cannot be distinguished based on the uncertainty on the log-evidences.  Those are the circular orbit with reflection, boosting, and ellipsoidal variations ($\ln Z = 25\,974.0 \pm 0.5$), and the circular orbit with reflection, boosting, ellipsoidal variations, and thermal emission ($\ln Z = 25\,974.0 \pm 0.4$).  In the case of circular orbits, the signals corresponding to reflection and thermal emission are both sinusoids and thus cannot be distinguished.  For this reason it is expected that adding thermal emissions into the model will not increase the log-evidence.

\begin{table}[h] 
\centering
\begin{tabular}{l c c}
			Model		&  Circular	&  Eccentric  \\
\hline
Refl. Only				& $25\,960.0 \pm 0.3 (5)$	& $25\;968.0 \pm 2.3 (7)$ \\

Boost Only			& $25\,954.0 \pm 1.8 (5)$    & $25\,962.0 \pm 2.0 (7)$\\

Ellips. Only			& $25\,799.0 \pm 0.5 (5)$ & $25\,968.0 \pm 1.0 (7)$\\			

Refl. + Boost. 			& $25\,960.0 \pm 0.4 (6)$ 	& $25\,970.0 \pm 0.8 (8)$ \\

Refl. + Ellips.			& $25\,971.0 \pm 0.4 (6)$	& $25\,971.0 \pm 0.85 (8)$ \\

Boost. + Ellips. 		& $25\,827.0 \pm 1.3 (5)$	& $25\,968.0 \pm 0.84 (7)$  \\

Refl. + Boost. + Ellips.	& $\bf{25\,974.0 \pm 0.5} (6)$	& $25\,972.0 \pm 1.2 (8)$ \\

Therm. + Boost + Ellipse. & $25\,973.0 \pm 0.4 (8)$ & $25\,970.0 \pm 1.7 (10)$ \\

Refl. + Boost. + Ellips. + Therm. & $\bf{25\,974.0 \pm 0.4} (8)$ & $25\,971.0 \pm 1.0 (10)$ \\

\hline
 Null & \multicolumn{2}{c}{$25\,774.0 \pm 1.1 (1)$} \\
\end{tabular}
\caption{MultiNest log evidences for 18 different models applied to the out-of-transit photometric signal of HAT-P-7b.  The model most favored to describe the out-of-transit data is in bold and the number of model parameters for each model are given in parentheses.}
\label{tbl:evidences}
\end{table}

As a by-product of MultiNest, one can perform posterior inferences on each model parameter to obtain means and standard deviations.  Table \ref{tbl:parameter estimates} shows the parameter estimates from one of the most favored out-of-transit models, in addition to the model that includes all photometric effects while taking transits and secondary eclipses into account. Model parameters include the stellar mass ($M_\star$) and radius ($R_\star$), planetary mass ($M_p$), radius ($R_p$), geometric albedo ($A_g$),  day ($T_d$) and night-side temperatures ($T_n$), and the intial mean anomaly ($M_0$), which represents the phase of the planet at the time of first observation. 

\begin{table}[h!]  
\centering
\small
\begin{tabular}{c c c c cc }
& Circular (R-B-E-T) & Circular (R-B-E-T) + Transits &  &   \\
 \hline
 Parameter  & Mean   & Mean   & \bf{Accepted}  \\
\hline
$M_{\star}$ $[M_{\odot}]$	&... 						&$1.471 \pm 0.001 M_\odot$		&	$\bf{1.43^{+0.08}_{-0.05} M_{\odot}}^a$	\\
$R_\star$  $[R_{\odot}]$		&...						&$2.0695 \pm 0.0005 R_\odot$		&	$\bf{1.84^{+0.23}_{-0.11} R_{\odot}}^a$	\\
 $M_0$ $[rad]$	 		&$4.77 \pm 0.04 rad$			&$4.9705 \pm 0.0001 rad$		&$...$	\\
i 	$[deg]$			&$69.47^o \pm 2.43^o$			&$81.4^o\,^{+0.01}_{-0.02}$	&	$\bf{84.1^o \pm 2.0^o}^a$	\\
$A_g$			&   $0.53 \pm 0.23$		& $0.088 \pm 0.026$		&	$...$	\\
$R_p$ 	$[R_J]$	&$0.83 \pm 0.30 R_J$			&$1.634 \pm 0.001 R_J$		&	$\bf{1.36^{+0.20}_{0.10}} R_J\,^a$	\\
$M_p$ 	$[M_J]$		&$1.67 \pm 0.64 M_J$			&$1.66 \pm 0.16 M_J$			&	$\bf{1.78^{+0.08}_{-0.05}} M_J\,^a$\\	
$T_{d}$ 	$[K]$		& $2633.00 \pm 1100.00 K$	&$2859.0 \pm 33.0 K$		&	$\bf{2730^{+150}_{-100}}K\,^b $\\
$T_n$	$[K]$		&	...					&$1332.0 \pm 756.0 K$		&	$...$\\

$\sigma$ (ppm)	& $28.00 \pm 0.23$		&$ 40.0 \pm 1.0$			&	$...$	\\
\hline
$\ln Z$ 			& $25\,974 \pm 0.4 $ 		&  $33\,238.0 \pm 0.9 $ &			

\end{tabular}
\caption{MultiNest parameter estimates for HAT-P-7b. Both models describe circular orbits and include reflected light, Doppler boosting, ellipsoidal variations and thermal emission (R-B-E-T).  Units for each parameter are listed in square brackets. $M_\odot$ and $R_\odot$ represent one solar mass and radius, respectively; whereas $M_J$ and $R_J$ refer to one Jupiter mass and radius, respectively.  The second column labeled "Circular (R-B-E-T)", displays the parameter estimates and log-evidence for the circular model applied to the out-of-transit data.  The third column, labeled "Circular (R-B-E-T) + Transits'', augments the circular model by including transits and secondary eclipses in addition to the other effects.  The bottom row lists the log-evidence, $\ln Z$, for both models.  Note that these two log-evidence values are not directly comparable as they are from datasets with a different number of datapoints.  Accepted values were taken from \citep{Pal+etal:2008}(a), and \citep{Esteves+etal:2013}(b).}
 \label{tbl:parameter estimates}
\end{table}

\begin{figure} [h!]
\includegraphics[ width= 15cm]{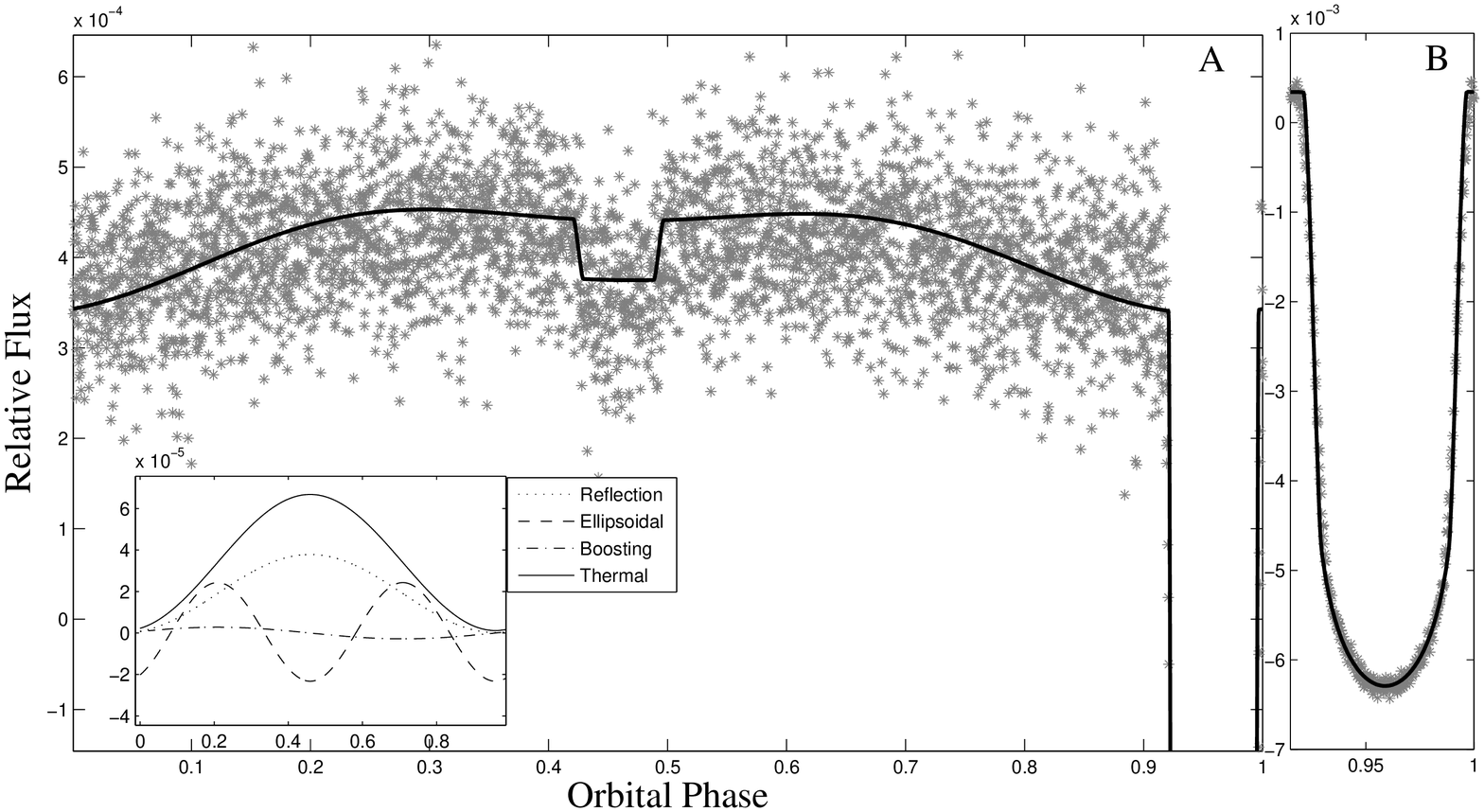}
\caption{EXONEST fit to HAT-P-7b data including both photometric effects and transits. The phase curve is shown with all four photometric effects in (A), and the primary transit is shown rescaled in (B).  The transit and secondary eclipse occur at orbital phases of 0.96 and 0.46, respectively. }
\label{fig:Kepler2b_transits}
\end{figure}

One can see that in the case of the out-of-transit model (2nd Column) the algorithm had difficulty estimating multiple parameters like the geometric albedo $A_g = 0.53 \pm 0.23$, the planetary radius $R_p = 0.83 \pm 0.30 R_J$, and the orbital inclination $i = 69.47^o \pm 2.43^o$; whereas when transits and secondary eclipses were taken into account these parameters were estimated much more accurately.  The low geometric albedo ($A_g = 0.088 \pm 0.026$) is consistent with short-period hot-Jupiters of this type \citep{Kane&Gelino:2010}, and the planetary mass and radius are within $1-$  and $2-\sigma$ of the accepted values, respectively.  An additional advantage to taking transits and secondary eclipses into account is that it allows one to estimate the day-side temperature of the planet to a higher degree of precision, and also allows one to obtain an estimate of the night-side temperature.  This is due to the fact that secondary eclipse depth is proportional to the day-side temperature of the planet, and the night-side temperature acts to increase the flux during transit, causing less of a dimming event than one would expect were there no thermal emissions from the night-side.  Figure \ref{fig:Kepler2b_transits} shows the transit model fit as well as the component photometric effects that comprise the out-of-transit signal.  Note that the secondary eclipse occurs at an orbital phase of approximately 0.46, and the transit occurs at an orbital phase of 0.96 (shown separately in Figure \ref{fig:Kepler2b_transits}B). 

\section{Conclusions}
The EXONEST algorithm has been employed to characterize the transiting exoplanet HAT-P-7b.  In the case of the out-of-transit data, based on the log-evidence values the two most favorable models to describe the data were circular orbits with reflection, boosting, and ellipsoidal variations, and those same three effects plus thermal emission.  This result is consistent with previous work as HAT-P-7b is expected to be in a circularized orbit about it's host star \citep{Pal+etal:2008}.  Since the planet is in a circular orbit, it is expected that distinguishing reflected and thermal light is impossible, which is corroborated by the log-evidence values.  EXONEST, which is now outfitted with a model that includes transits and secondary eclipses,  in addition to the four other photometric effects, was used to compare the parameter estimates between the models that exclude transits to those from the model that includes transits.  Based on the model that includes transits and secondary eclipses, HAT-P-7b is a very dark planet, reflecting only $\sim9\%$ of the incident star-light.  It is also extremely hot with a day-side temperature of $T_d = 2859 \pm 33K$ consistent with the estimate obtained by Esteves et al. 2013.  The planetary radius and orbital inclination, which were not well estimated in the case of out-of-transit data, had very small uncertainties associated with them in the case of transits and secondary eclipses.  However, these two parameters were slighty outside the $1-\sigma$ range of previously published values.  

Bayesian model selection gives one the opportunity to effectively characterize exoplanets by considering the photometric effects detectable in Kepler light curves.  This is ever more important as both the instrumental precision of telescopes, and the amount of photometric data continue to increase.





\bibliographystyle{aipproc}   

\bibliography{bibliography.bib}


\end{document}